\documentclass[review]{elsarticle}
\usepackage{graphics}
\usepackage{graphicx}
\usepackage{float} % Place images here with [H]
\usepackage{bm}
\usepackage[nomarkers,notablist,nofiglist]{endfloat}
\usepackage{subfigure}
\usepackage{lineno}
\modulolinenumbers[5]
\usepackage{siunitx}         
\graphicspath{ {./Figures/} } 
\usepackage{soul}
\usepackage{miller}
\usepackage{amsmath}
\usepackage{geometry} % to change the page dimensions
\usepackage{color} 
\usepackage{outlines}
\usepackage{upgreek}
\DeclareSIUnit\angstrom{\text {Å}}

\journal{arXiv}

\begin{document}

\begin{frontmatter}
	
    \title{The influence of substantial intragranular orientation gradients on the micromechanical response of heavily-worked material}
    
    \author[1,2]{Karthik Shankar\corref{cor1}}
    \author[3]{Meddelin Setiawan\corref{cor1}}
    \author[4]{Katherine S. Shanks}
    \author[5]{Matthew E. Krug}
    \author[1]{Matthew P. Kasemer\corref{cor2}}
        \ead{mkasemer@eng.ua.edu}
    \author[3]{Darren C. Pagan\corref{cor2}}
        \ead{dcp5303@psu.edu}
    \cortext[cor1]{Co-First Author}
    \cortext[cor2]{Corresponding Author}
	
    \address[1]{Department of Mechanical Engineering, The University of Alabama, Tuscaloosa, AL 35487}
    \address[2]{Structures, Safety, and Airworthiness, The Boeing Company, Long Beach, CA 90808}
    \address[3]{Materials Science and Engineering, The Pennsylvania State University, University Park, PA 16802}
    \address[4]{Cornell High Energy Synchrotron Source, Cornell University, Ithaca, NY 14853}
    \address[5]{Air Force Research Laboratory, Materials and Manufacturing Directorate, Wright-Patterson Air Force Base, OH 45433}
    
\begin{abstract}

In this study, we employ high-energy X-ray characterization to examine the role of relatively large amounts of intragranular lattice misorientation -- present after many thermomechanical processes -- on the micromechanical response of Al-7085 with a modified T7452 temper. We utilize near-field high energy X-ray diffraction microscopy (HEDM) to measure three-dimensional (3D) spatial orientation fields, facilitated by a novel method that utilizes grain orientation envelopes measured using far-field HEDM to enable reconstruction of grains with intragranular orientation spreads greater than \SI{20}{\degree}. We then assess the consequences of consideration of intragranular orientation fields on the predicted deformation response of the sample through  3D micromechanical simulations of the forged Al-7085. We construct two virtual polycrystalline specimens for use in simulations: the first a faithful representation of the HEDM reconstruction, the second a microstructure with no intragranular misorientation (i.e., grain-averaged orientations). We find significant differences in the predicted deformation mechanism activation, distribution of stress, and distribution of plastic strain between simulations containing intragranular misorientation and those with grain-averaged orientations, indicating the necessity for consideration of intragranular orientation fields for accurate predictions. Further, the influence of elastic anisotropy is discussed, along with the effects of intragranular misorientation on fatigue life through the calculation and analysis of fatigue indicator parameters.

\end{abstract}

\end{frontmatter}

%\linenumbers

\section{Introduction}

In addition to grain reorientation and macroscopic texture evolution that occurs during thermomechanical processing, there is generally an accompanying development of crystallographic misorientation within individual grains~\cite{hirsch1952study,hansen2001new}. Similarly, intragranular misorientation develops during many solidification processes~\cite{doherty1984microstructure,wang1990characterization}, including during most additive manufacturing~\cite{yoo2018identifying,voisin2021new}. The magnitudes of intragranular misorientation present have the potential to be significant, on the order of 10's of degrees. As local orientation is highly influential on the micromechanical response, heterogeneous intragranular orientation fields have the potential to lead to large variations in mechanical state and active plastic deformation mechanisms. However, the challenge with micromechanical modeling---and ultimately quantifying---these effects lies with the ability to instantiate a model with realistic intragranular orientation fields (i.e., including intragranular misorientation). This misorientation present within grains results from underlying three-dimensional (3D) geometrically necessary dislocation (or Nye) tensor fields, which are described by a very large parameter space that is not readily evaluated. A more tractable approach is to instantiate simulations with experimentally measured orientation fields which, while not capturing the full parameter space, do reflect realistic microstructural conditions. However, means to collect 3D orientation fields, particularly in heavily deformed materials, are limited and often destructive~\cite{echlin2012new}, the latter eliminating possibility of further {\it in situ} micromechanical experiments on the measured sample. To address this issue, we present a novel methodology to non-destructively reconstruct 3D orientation fields with large intragranular orientation gradients using high-energy X-ray diffraction microscopy (HEDM), and demonstrate this methodology on a forged Al-7085 sample. The data is used to instantiate crystal plasticity finite element method (CPFEM) simulations and explore what role orientation fields with appreciable intragranular misorientation play in (micro)mechanical response.

Electron backscatter diffraction (EBSD) has emerged as the dominant method to characterize orientations within metallic alloys, in addition to more recent but less-employed optical and ultrasonic measurements~\cite{Jin2020,vanWees2023,He2022}. These measurements are two-dimensional (2D) and restricted to the sample surface, which while powerful in providing insight into the nature of orientation fields in processed and deformed alloys, do not allow access to the full 3D orientation field. Combining these methods with serial sectioning, particularly EBSD~\cite{echlin2012new,echlin2020serial}, provides the opportunity to fully map orientation fields in 3D, but the method is destructive, preventing future mechanical testing, thus disallowing the direct experimental establishment of structure-property relationships. In contrast, 3D X-ray methods such as near-field HEDM (nf-HEDM)~\cite{suter2006forward,nygren2020algorithm} and diffraction contrast tomography~\cite{johnson2008x} can non-destructively map comparable volumes, albeit with reduced spatial resolution (approximately 1-\SI{5}{\micro\meter}) and (with standard reconstruction algorithms) limitations on the magnitude of intragranular orientation that can be reconstructed (approximately 3-\SI{5}{\degree}) due to the large space of possible orientation configurations that need to be tested. These X-ray data have been used to instantiate both FEM~\cite{proudhon2016coupling,turner2017crystal,renversade2024intra} and Fast-Fourier Transform~\cite{pokharel2014polycrystal,pokharel2017instantiation,tari2018validation} micromechanical simulations from well-annealed alloy systems with low-defect content and minimal intragranular misorientation. However, a recently developed nf-HEDM reconstruction method utilizes complimentary orientation distribution information measured in the far-field (ff-HEDM) to constrain the number of orientation configurations that need to be tested~\cite{nygren2020algorithm}. As we will demonstrate in this study, this effectively increases the magnitude of intragranular orientation distributions that can be successfully reconstructed, providing a new class of data that can be used for micromechanical simulation instantiation.

We present a novel methodology to reconstruct the spatial fields of orientations in 3D samples with large orientation gradients using a combined nf-HEDM and ff-HEDM approach, and further explore how large orientation gradients that arise during metal processing influence micromechanical response via a suite of crystal plasticity simulations. First, we describe the experimental and novel reconstruction methodology used to reconstruct 3D orientation fields with the presence of appreciable intragranular orientation gradients in Al-7085. Next, we describe the instantiation of the virtual polycrystals for use in crystal plasticity simulations, in particular the instantiation of the faithful representation of the experimentally-deduced polycrystal, as well as the grain-averaged homogenized polycrystal. Next, results comparing the predicted micromechanical response of these two samples are presented. Lastly, we discuss the relationships between intragranular orientation gradients and plastic deformation mechanism activation, the influence of elastic anisotropy, and finally the fatigue response. 

Herein, an overbar indicates an averaged or macroscopic quantity while a tilde indicates an equivalent scalar of a second-rank tensor.

\section{Material and 3D Microstructure Measurements}

\subsection{Material and Measurement Description}

\begin{figure}[h]
      \centering \includegraphics[width=0.7\textwidth]{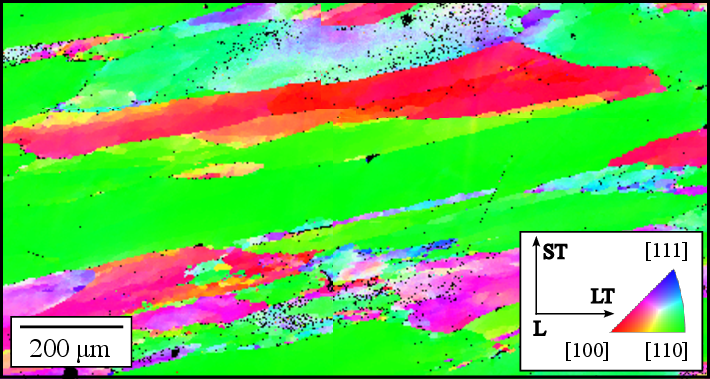}
      %\vspace{-0.15in}
      \caption{Electron backscatter diffraction (EBSD) image of the Al-7085 material tested and modeled. The short transverse (ST) and the long transverse (LT) directions are in the image plane and longitudinal (L) is out of the plane. Grains are colored by inverse pole figure with respect to the ST direction.}
      \label{fig:ebsd}
\end{figure}

The material used in this study is based on hand-forged Al-7085-T7452, which is used extensively in airframe parts. Prior to characterization, we altered the temper by subjecting the material to additional heat treatment in an effort to promote a moderate increase in grain size while retaining the appreciable intragranular misorientation present from forged processing (i.e., the heat treating was chosen to avoid recrystallization). Specifically, we heat treated the material at \SI{400}{\celsius} for \SI{30}{\minute} after the forging process. We present an EBSD image containing the long transverse (LT) and short transverse (ST) directions of the forging in Figure \ref{fig:ebsd}. We can see that the grain size is significantly smaller in the ST direction, as expected. Even after heat treatment, the specimen is still heavily textured and we observe appreciable orientation spread through the grains. We note that several intermetallic phases \cite{li2011identification} can be found in Al-7085, but the size of these phases is generally below the resolution of both the measurement and modeling methodologies employed in this study.

To characterize the microstructure of the sample in 3D, we collected X-ray data at the Forming and Shaping Technology (FAST) Beamline at the Cornell High Energy Synchrotron Source (CHESS). In total, we collected far-field (ff-HEDM), near-field (nf-HEDM) HEDM, and X-ray computed tomography (XCT) measurements of the specimen. We utilized the RAMS2 system~\cite{shade2015rotational} to leverage its robust integration with combined nf-HEDM and ff-HEDM measurement functionality. Consequently, a specimen compatible with mounting in the RAMS2 load frame was extracted from the larger forged material. The specimen was cut with a gauge cross section of \SI{1}{\milli\meter} $\times$ \SI{1}{\milli\meter}, and such that the ST forging direction was aligned with the specimen long axis, while the LT and L directions aligned with the sample transverse face normals.

We utilized an incoming X-ray beam with an energy of \SI{41.991}{\kilo\electronvolt} (Pr K$_\alpha$-edge) and a width of \SI{2}{\milli\meter} to fully illuminate the sample cross section. We utilized a beam height of \SI{0.135}{\milli\meter}, and we probed 4 volumes with a spacing of \SI{0.125}{\milli\meter} (the vertically oversized beam facilitates stitching and allows data at the beam edges to be discarded). In total, we probed a volume spanning \SI{1}{\milli\meter} along L,  \SI{1}{\milli\meter} along LT, and \SI{0.5}{\milli\meter} along ST. For both nf-HEDM and ff-HEDM, we collected diffraction data across a full \SI{360}{\degree} in \SI{0.25}{\degree} increments (1440 images total) as we continuously rotated the specimen about the ST direction. Prior to the HEDM measurements, we performed an XCT scan using a \SI{2}{\milli\meter} wide by \SI{1}{\milli\meter} tall beam with X-ray absorption radiographs made at \SI{0.5}{\degree} increments over \SI{360}{\degree}. These XCT measurements facilitate the establishment of the physical extent of the specimen to speed data reconstruction. 

Schematics of the instrument geometry can be found in more detail in~\cite{nygren2020cartography}. We measured the far-field HEDM data using 2 Dexela 2923 area detectors (3888 $\times$ 3072 pixels and \SI{74.8}{\micro\meter} pixel size). We placed these detectors \SI{751}{\milli\meter} behind the specimen, and they were arrayed side-by-side with an approximately \SI{2}{\centi\meter} gap between them through which the incoming beam passed. At this distance, 6 complete rings were present across the two detectors. We calibrated the far-field detector positions using a two-step process with a CeO$_2$ powder calibrant to find rough instrument parameters, followed by refinement using a NIST single crystal ruby standard. We collected the nf-HEDM images on a Retiga optical detector coupled to a LuAg:Ce scintillator with a 5$\times$ objective lens, providing an effective pixel pitch of \SI{1.48}{\micro\meter}. For nf-HEDM measurements, we placed the detector \SI{6.85}{\milli\meter} away from the specimen. The near field instrument geometry was determined as part of the reconstruction process, as instrument parameters are selected which maximize the quality (completeness, see below) of the reconstruction.

\subsection{Data Processing and Microstructure Reconstructions}

To reconstruct the spatial orientation field, we primarily follow the methods described in~\cite{bucsek2019three,nygren2020algorithm}. The major difference in this study is that we test large regions of orientation space (similarly to that described in Long and Miller~\cite{long2019statistical}), as opposed to the method of testing small clusters of orientations around grain-averaged orientations found using ff-HEDM~\cite{bernier2011far}. The first step of this process is to determine the orientations (and intragranular orientation spreads) that are present within the volume probed. We achieve this via a modified ff-HEDM indexing procedure. Typically, far-field indexing consists of testing a series of orientations by forward projecting where diffracted intensity would appear on a far-field detector(s) as a specimen is rotated. The positions of these intensities are then compared to those measured experimentally. We determine a completeness metric for the tested orientations defined as the percentage of predicted diffracted intensities found in the data, and if above a certain threshold, the orientation is determined to be present. Usually, the list of trial orientations are those along crystallographic fibers determined from the peak positions. Here, we instead test a list of trial orientations that fully span the fundamental region of orientation space for cubic crystal symmetry. Specifically, we discretize the fundamental region with a regular grid of orientations spaced \SI{0.5}{\degree} apart, which leads to approximately 5.2 million discrete trial orientations. We note that the spacing chosen for these test orientations ultimately dictates the orientation resolution of the reconstructed spatial orientation field.

\begin{figure}[h]
      \centering \includegraphics[width=1.0\textwidth]{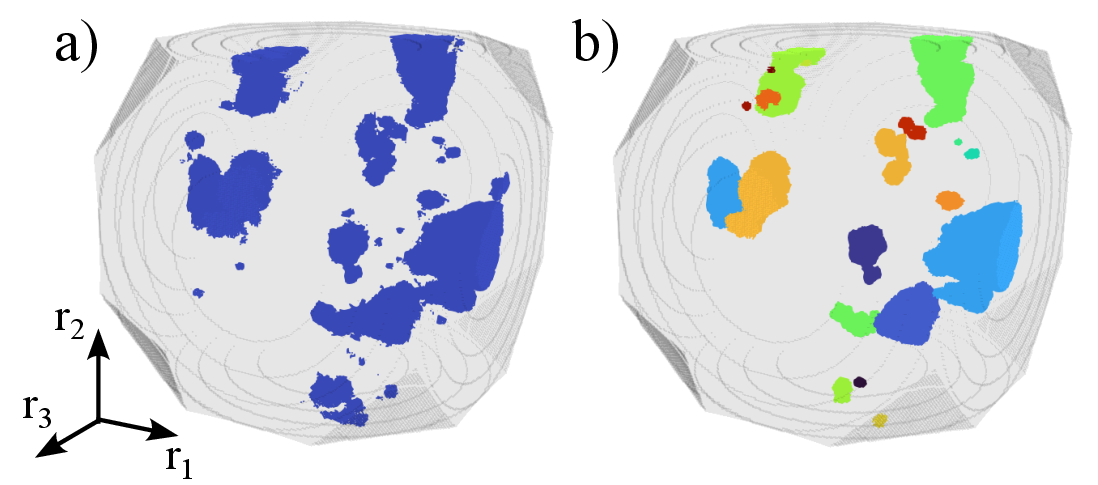}
      %\vspace{-0.15in}
      \caption{a) Orientations measured within the full diffraction volume, represented in the Rodrigues parameterization and depicted in the cubic fundamental region of Rodrigues space, and b) grain orientation envelopes (GOEs) found using the DBSCAN clustering algorithm, excluding de-noised points associated with isolated, minor volumes.}
      \label{fig:blobs}
\end{figure}

Following this methodology, we perform indexing on the far-field diffraction data using the trial orientation grid. We employ a completeness threshold of 0.79, selected from direct analysis of the full distribution of completeness values (i.e., we selected a threshold necessary to index the majority of the material volume without incorporating spurious orientation solutions). We utilize a lattice parameter ($a_0$) of \SI{4.056}{\angstrom} for predicting diffracted intensity positions. After indexing all four volumes probed during measurement, we concatenate the lists of individual orientations from each volume into a global list for orientation testing. In total, we find 181,300 orientations in the full diffraction volume, and present a plot of these orientations in the cubic symmetry fundamental region of Rodrigues space (using the Rodrigues orientation parameterization, $\bm{r}$, and a crystal-to-sample convention) in Figure~\ref{fig:blobs}a. We observe distinct, closed ``envelopes'' of orientations (in other words: contiguous point-clouds), which we will herein refer to as grain orientation envelopes (GOEs). We note that these GOEs generally correspond to grains within the diffraction volume (i.e., the orientations within GOEs are contiguous in real-space), indicating that they do indeed belong to single grains. To isolate these different envelopes, we employ the DBSCAN clustering algorithm~\cite{ester1996density} with parameters of $\epsilon=0.02$ (the clustering distance) and 32 orientations for the minimum cluster size, with parameters chosen to isolate envelopes that were visually apparent by inspection. After clustering, we manually identified envelopes which are discontinuous in orientation space due to crystal symmetry (i.e., exit one side and appear on the other, see~\cite{kumardawsonneoeulerian1998}) and combine them. The GOEs which reside on symmetric equivalencies of the fundamental region boundaries tend to have similar point cloud shapes on the boundaries, lending confidence that they indeed belong to the same GOE and are separated only due to crystal symmetry. We find 17 distinct GOEs, as shown in Figure~\ref{fig:blobs}b (colored by GOE), and note the presence of significant variation in GOE size in orientation space, which is particularly noteworthy as several of these GOEs indicate granular orientation spreads on the order of \SI{10}{\degree} to \SI{20}{\degree}.

\begin{figure}[h]
      \centering \includegraphics[width=1.0\textwidth]{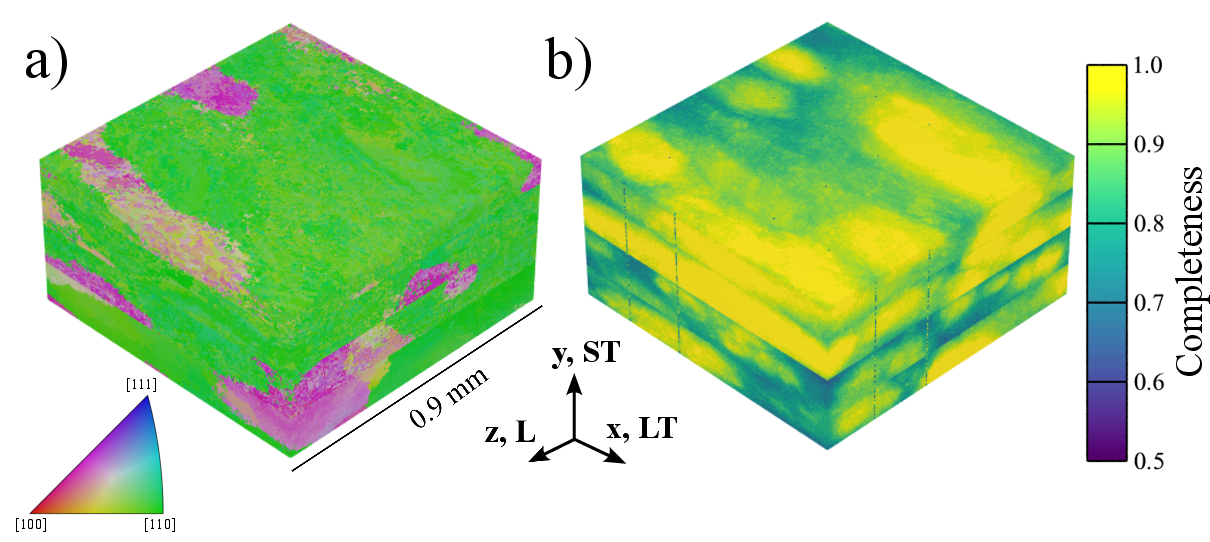}
      %\vspace{-0.15in}
      \caption{a) 3D reconstructed spatial orientation field of the forged Al-7085 specimen plotted on an inverse pole figure color scale with respect to $\bm{y}$, the loading direction, and b) reconstruction completeness (predicted vs. measured diffracted intensity) within the same volume. The relationship between the coordinate system and the forging directions short transverse (ST), long transverse (LT), and longitudinal (L) are also provided.}
      \label{fig:nf}
\end{figure}

We next utilize the orientations within the GOEs (Figure~\ref{fig:blobs}a) as input trial orientations for the nf-HEDM algorithm. As previously mentioned, our approach primarily follows that reported in~\cite{nygren2020algorithm} in which a list of trial orientations is tested at every real-space voxel of interest. First, we repeatedly reconstruct a single layer from the diffraction volume (established previously using XCT data) with a coarse spatial resolution (\SI{10}{\micro\meter} voxel spacing), while iteratively varying the instrument parameters: the sample to detector distance, intersection of the beam with the detector, and tilts of the detector face. We repeat the process to increase the ratio of predicted diffracted intensity to measure diffracted intensity (i.e., completeness) and the sharpness of grain boundaries within the reconstructed layer (analogous to focusing a lens). With optimal instrument parameters, we reconstruct the four nf-HEDM diffraction volumes using a finer voxel spacing of \SI{5}{\micro\meter}, chosen to approximately match the instrument resolution, and finally merge the separate nf-HEDM volumes in post-processing. We present the nf-HEDM reconstruction in Figure~\ref{fig:nf}a, as well as the corresponding completeness of the reconstruction, which varies from 0.7 to 1.0, in Figure~\ref{fig:nf}b. We note that there is a relatively strong crystallographic texture with a significant volume fraction of material near the \hkl[1 1 0] crystallographic direction along ST (also apparent in Figure~\ref{fig:blobs}a).

\section{Crystal Plasticity Finite Element Modeling}

\subsection{Crystal Plasticity Model and Material Parameters}

We perform micromechanical simulations using the FEPX CPFEM package. Here we provide only a truncated description of the model employed in FEPX below. A complete implementation is detailed in~\cite{marin1998modelling,marin1998elastoplastic,dawson2015fepx,fepx}. The FEPX framework considers ductile, isothermal, quasi-static material behavior, and employs a non-linear finite element solver to approximate the solution to the equilibrium via an implicit formulation to force the reduction of a global weighted residual (thus ensuring the approximate enforcement of stress equilibrium). In this formulation, the lattice orientation and slip system strength are stored as evolving state variables at the quadrature points of each element.

We describe the kinematics of material motion via a large strain formulation in which the deformation gradient is multiplicatively decomposed into an elastic portion, a rotation (material and lattice), and a plastic portion. To consider the elastic response of the material, we employ the anisotropic form of Hooke's law:
\begin{equation}
    \label{eq:hooke}
    \bm{\sigma}=\underline{\bm{C}}(\bm{r}):\bm{\varepsilon_E}
    \quad ,
\end{equation}
where $\bm{\sigma}$ is the stress tensor, $\underline{\bm{C}}$ is the lattice-orientation-dependent elastic stiffness tensor, $\bm{r}$ is the lattice orientation, and $\bm{\varepsilon_E}$ is the elastic strain tensor. Here the stiffness tensor has three independent constants ($C^C_{11}$ , $C^C_{12}$ , and $C^C_{44}$ in Voigt notation in the crystal frame) due to the cubic symmetry of Al-7085. We employ the stiffness constants used for simulations in~\cite{zhang2022effect}, which are summarized in Table~\ref{tab:elastic_properties}.

\begin{table}[h!]
    \centering
    \begin{tabular}{c|c c c c}
    Material & {\bf $C_{11}$ (\SI{}{\giga\pascal})} & {\bf $C_{12}$ (\SI{}{\giga\pascal})} & {\bf $C_{44}$ (\SI{}{\giga\pascal})} & {\bf $A$ (-)} \\
    \hline
    Isotropic & 111.2 & 57.4 & 26.4 & 0.94 \\
    Anisotropic & 111.2 & 57.4 & 79.2 & 2.98 \\
    \end{tabular}
    \caption{Single crystal elastic moduli ($C^C_{11}$ , $C^C_{12}$ , and $C^C_{44}$ in Voigt notation) and associated Zener ratios $A$ utilized in the crystal plasticity simulations. The ``anisotropic'' moduli are chosen such that the bulk modulus remains the same as the ``isotropic'' moduli, but with an increase in Zener ratio.}
    \label{tab:elastic_properties}
\end{table}

We model plasticity via a rate-dependent slip kinetics formulation, restricting slip to a specific set of slip modes (in the case of FCC Al-7085, we assume slip is restricted to the 12, \hkl<1 1 0>\hkl{1 1 1}-type systems). In this formulation, the plastic velocity gradient $\bm{L}_P$ is described as a sum of the shearing rates on the slip systems, $\alpha$:
\begin{equation}
    \label{eq:lp}
    \bm{L}_P=\sum_\alpha{\dot{\gamma}_\alpha (\bm{s_\alpha}\otimes\bm{n_\alpha})}
    \quad ,
\end{equation}
where $\dot{\gamma}_\alpha$ is the slip system shearing rate, $\bm{s_\alpha}$ is the slip direction, and $\bm{n_\alpha}$ is the slip plane normal. Following standard constitutive assumptions, the plastic velocity gradient is able to deform the material, but leaves the lattice unchanged. We model slip kinetics on a per-system basis via:
\begin{equation}
    \label{eq:gammadot}
    \dot{\gamma}_\alpha=\dot{\gamma}_0  
    \frac{\tau_\alpha}{\tau^*}
    \left| 
    \frac{\tau_\alpha}{\tau^*}
    \right|^{1/m-1}
    \quad ,
\end{equation}
where $\tau_\alpha$ is the stress projected onto slip system (i.e., the resolved shear stress), $\tau^*$ is the current slip system strength (or hardness), and $m$ is the power law exponent. We calculate the resovled shear stress on each slip system as:
\begin{equation}
	\tau_\alpha=\bm{\sigma}:(\bm{s_\alpha}\otimes\bm{n_\alpha})=\bm{\sigma'}:(\bm{s_\alpha}\otimes\bm{n_\alpha})
	\quad ,
\end{equation}
where $\bm{\sigma'}$ is the deviatoric portion of the stress. We allow the current slip system strength, $\tau^*$, to evolve as a function of local plasticity. We employ an isotropic (i.e., the single crystal yield surface may dilate, but retains its shape in deviatoric stress space) saturation-style hardening model:
\begin{equation}
    \label{eq:hardening}
    \dot{\tau^*}=h_0  
    \frac{\tau_S-\tau^*}{\tau_S-\tau_0}\sum_\alpha |\dot{\gamma}_\alpha|
    \quad ,
\end{equation}
where $\tau_0$ is the initial slip system strength, $\tau_S$ is the slip system strength at saturation, and $h_0$ is the fixed-state hardening coefficient. We optimize the plasticity parameters by fitting to large-strain macroscopic stress-strain data measured on Al-7085 in~\cite{zhang2022effect}, the results of which are summarized in Table~\ref{tab:plastic_properties}.

\begin{table}[h!]
    \centering
    \begin{tabular}{c c c c c c}
    $m$ & $\dot{\gamma}_0$ {\bf (\SI{}{1\per\second)}} & $h_0$ {\bf (\SI{}{\mega\pascal})} & $\tau_{0}$ {\bf (\SI{}{\mega\pascal})} & $\tau_s$ {\bf (\SI{}{\mega\pascal})} \\
    \hline
    0.05 & 0.001 & 300 & 130 & 230 \\
    \end{tabular}
    \caption{Crystal plasticity modeling parameters utilized for Al-7085.}
    \label{tab:plastic_properties}
\end{table}

To aid in analysis of the predicted deformation fields, we will analyze the total local plastic strain, $\tilde{\varepsilon}_P$, which we define as:
\begin{equation}
    \label{eq:ep}
    \tilde{\varepsilon}_P=\int{\tilde{D}_{P} dt}
    \quad ,
\end{equation} 
where $\tilde{D}_{P}$ is the effective plastic deformation rate:
\begin{equation}
    \tilde{D}_{P}  = \sqrt{\frac{2}{3}\bm{D}_P : \bm{D}_P}
    \quad ,
\end{equation}
and the plastic deformation rate tensor, $\bm{D}_P$, is the symmetric portion of the plastic velocity gradient (Eq.~\ref{eq:lp}).

\subsection{Virtual Microstructure Instantiation and Loading}

Overall, we choose to instantiate two separate virtual microstructures for use in the CPFE simulations in an effort to determine the effects of incorporating appreciable intragranular orientation spreads: the first, a faithful representation of the spatial orientation field, and the second, a structure with grain-averaged orientations applied homogeneously within each grain. In both structures, we utilize the nf-HEDM reconstruction in Figure~\ref{fig:nf}a to instantiate the spatial location of the grains in the sample. In this way, we faithfully represent the geometric morphology of the granular structure in the virtual samples. To achieve this, we construct a regular grid of voxels and perform a direct mapping of the measured orientations of the experimental microstructure to that of the virtual grid. This virtual grid may have coarser or finer resolution than that of the experimental grid, depending on desired model resolution. In this study, due to computational constraints, we utilize a virtual polycrystal with a coarser voxel grid than the experimental grid. In this case, we assign an orientation to a point in a coarse grid of the virtual polycrystal based on the dominant experimental orientation in its direct vicinity (i.e., the most prevalent orientation within the element bounds).

In the first of the two microstructures (i.e., a faithful representation of the spatial orientation field), we apply local orientations utilizing a similar scheme above employed for the assignment of grain ID. Here, a point in a coarse grid of the virtual polycrystal is assigned an orientation based on the average orientation of the experimental voxels with common grain ID in the local vicinity using orientation averaging functionality in the HEXRD package that accounts for crystal symmetry. In this way, we ensure that the orientation field of the virtual polycrystal matches that measured experimentally. In the second of the two microstructures, we assign a single orientation to each voxel with a common grain ID, where we calculate this orientation to be equal to the average of all orientations in the experimental dataset with the same grain ID (i.e., we average over the GOEs shown in Figure~\ref{fig:blobs}a). This method is consistent with the approaches of the overwhelming majority of crystal plasticity studies, and obscures the gradient of orientations present within grains. From this point onwards, we refer to the first of the two microstructures as the ``gradient'' structure, and the second of the two as the ``homogenized'' structure. In both the gradient and homogenized microstructures, each voxel in the grid is split into 6, 10-node tetrahedral elements, with each element assigned the same grain ID and orientation as its parent voxel.

\begin{figure}[h!]
    \centering \includegraphics[width=1.0\textwidth]{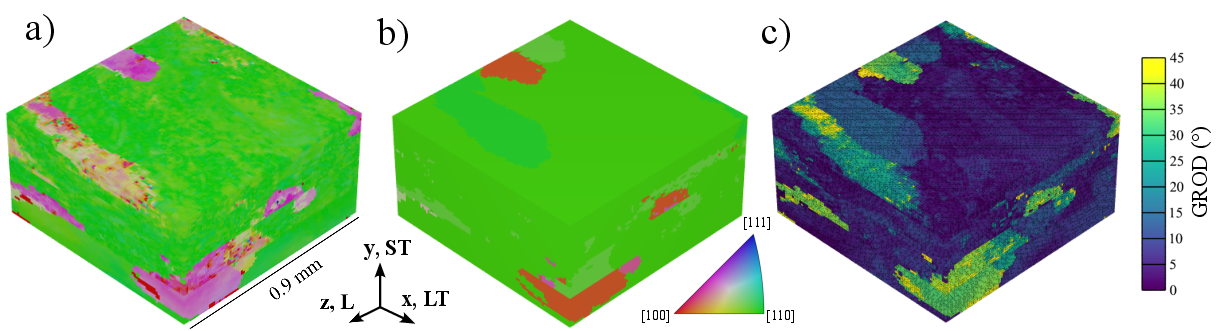}
    \caption{Virtual microstructure samples utilizing orientation fields from nf-HEDM, specifically a) a sample considering a faithful mapping of the orientations as measured via HEDM, where each grain has intragranular orientation gradients (i.e., the ``gradient'' virtual microstructure), and b) a sample where orientations within each grain are averaged over the grain orientation envelopes (GOEs) in orientation space (i.e., the ``homogenized'' virtual microstructure). The elements are colored on an inverse pole figure color scale with respect to $\bm{y}$, the loading direction. c) Grain reference orientation deviation (GROD) between the gradient and homogenized virtual microstructures.}
    \label{fig:sim_structure}
\end{figure}

Following the general procedure above, we begin by first removing a thin layer of material on the experimental data set associated with the rough sample surface. We reduce the sample size from 1 $\times$ 0.5 $\times$ \SI{1}{\milli\meter\cubed} to {0.9} $\times$ {0.45} $\times$  \SI{0.9}{\milli\meter\cubed} due to this truncation process (i.e., the reduced volume is left comprised of {180 $\times$ 90 $\times$ 180} voxels). We further perform de-noising of the nf-HEDM orientation field by finding isolated volumes of orientation with sizes less than 64 voxels and then setting these orientations to match the neighboring region (compare: Figure~\ref{fig:blobs}a and b, and note the removal of small orientation features as a consequence of this de-noising). We map the nf-HEDM microstructure to the virtual polycrystal grid with mesh coarsening of 2$\times$---i.e., we concatenate 8 experimental voxels (2 $\times$ 2 $\times$ 2) into a single voxel in the virtual polycrystal grid, resulting in a virtual polycrystal with a voxel grid of {90 $\times$ 45 $\times$ 90} voxels and thus reducing the mesh size 8$\times$ compared to a direct 1-to-1 map. Considering the 6 tetrahedral elements per voxel, the overall mesh contains approximately 2.2 million elements. We present the gradient and homogenized virtual microstructures in Figure~\ref{fig:sim_structure}a and Figure~\ref{fig:sim_structure}b respectively. The grain reference orientation deviation (GROD) between the two microstructures calculated on an element-by-element basis is shown in Figure~\ref{fig:sim_structure}c.

We apply loading to the two virtual microstructures along their ST ($\bm{y}$) directions to explore the variation of micromechanical response due to the differences in orientation present in the material. We load through the yield point and into fully-developed plastic flow to macroscopic engineering strains of 0.2. We employ boundary conditions via velocities applied at select nodes of the finite element mesh. We apply minimal boundary conditions to the sample, where the domain is constrained to apply axial deformation yet disallow rigid body rotation or motion, by applying axial velocities on two surfaces, and fixing two nodes in orthogonal directions. This, overall, allows the polycrystal to deform with minimal undue influence from the boundary conditions which may over-constrain portions of the domain and lead to stress concentrations and over-predictions. On the loading surface (normal along $\bm{y}$), we apply velocities on the control surface to enforce a constant engineering strain rate of \SI{0.001}{\per\second}.

\section{Simulation Results}

\subsection{Macroscopic Behavior}

\begin{figure}[h!]
    \centering
    \includegraphics[width=.6\textwidth]{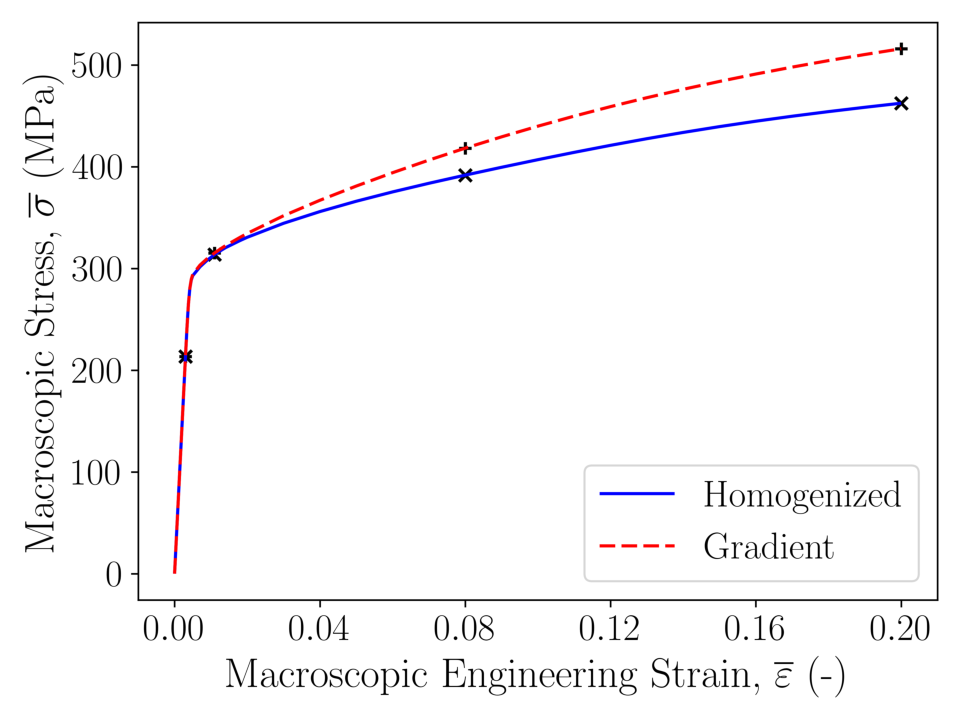}
    \caption{Macroscopic uniaxial stress-strain curves for the gradient and homogenized virtual microstructures loaded along the ST ($\bm{y}$) direction. Markers shown are at 0.003, 0.011, 0.080, 0.200 macroscopic engineering strain, in which more detailed element-scale analysis is performed.}
    \label{fig:stress-strain}
\end{figure}

In this section, we contrast the mechanical response of the gradient microstructure versus the homogenized microstructure. We present the macroscopic stress-strain ($\bar{\sigma}$-$\bar{\varepsilon}$) response up to engineering strains of 0.2 (20\%) for both microstructures in Figure~\ref{fig:stress-strain}. We provide more detailed analysis at 4 points along the loading history, marked in Figure~\ref{fig:stress-strain}, specifically at macroscopic engineering strains of 0.003 (elastic regime), 0.011 (yield), 0.08 (post-yield), and 0.2 (fully-developed flow). Again, we note that we utilize the same elasticity and plasticity parameters (first row of Table~\ref{tab:elastic_properties}, and Table~\ref{tab:plastic_properties}, respectively) when simulating both the gradient and homogenized microstructures---i.e., the simulations differ only in microtexture / spatial distributions of orientations. While the yield stresses of the two microstructures are nearly identical (see Table~\ref{tab:macro_response}), we observe that the flow stresses of the two virtual microstructures begin to deviate significantly from one another shortly after macroscopic yield and the differences continue to grow as plasticity develops (e.g., at $\bar{\varepsilon}=0.2$, the flow stress of the gradient microstructure is over \SI{50}{\mega\pascal} higher than that of the homogenized microstructure). The macroscopic behavior of the gradient microstructure exhibits an appreciably higher macroscopic hardening rate, and does not reach a point of macroscopic saturation (i.e., a peak in the stress-strain curve). Conversely, the macroscopic behavior of the homogenized microstructure exhibits a lower macroscopic hardening rate, and is approaching the peak of its stress-strain behavior. These differences in macroscopic strength and hardening rate are a result of the differences in local microtexture between the two virtual specimens rather than a size effect.

\begin{table}[h!]
    \centering
    \begin{tabular}{c|c c }
    Material & Homogenized &  Gradient\\
    \hline
    Isotropic & 293  & 295 \\
    Anisotropic & 304 & 304 \\
    \end{tabular}
    \caption{Yield strength, $\sigma_y$ {\bf (\SI{}{\mega\pascal})}, for simulations utilizing the ``isotropic''~\cite{zhang2022effect} and ``anisotropic'' elastic moduli calculated from monotonic tensile simulations using a 0.1\% offset method.}
    \label{tab:macro_response}
\end{table}

\subsection{Element-Scale Stress Behavior}

\begin{figure}[h!]
    \centering \includegraphics[width=0.7\textwidth]{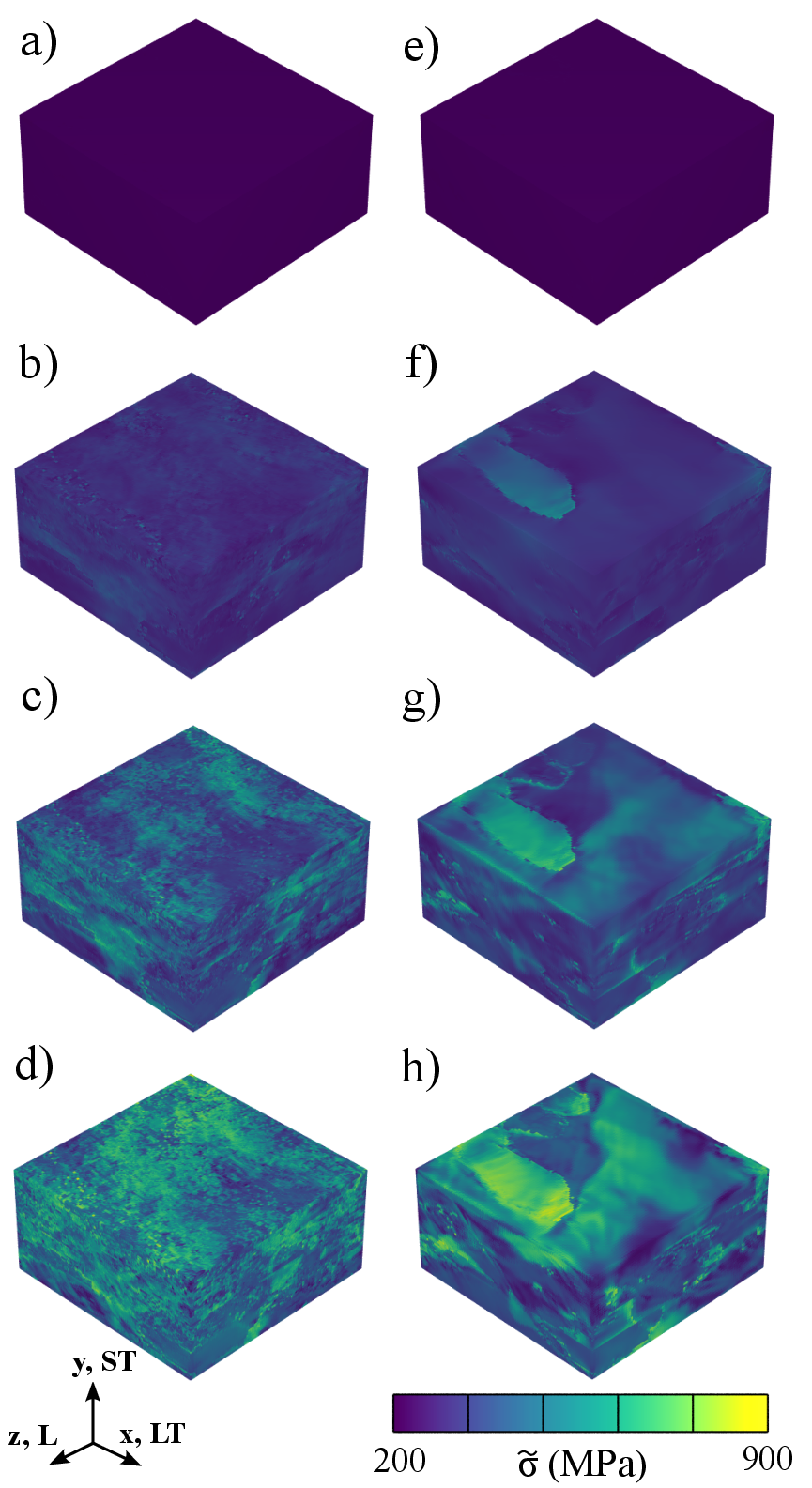}
    \caption{Distributions of von Mises stress, $\tilde{\sigma}$, at (a,e) 0.003,  (b,f) 0.011, (c,g) 0.080, and (d,h) 0.200 macroscopic engineering strain in the gradient (left) and homogenized (right) virtual microstructures.}
    \label{fig:von_Mises3d}
\end{figure}

Next, we can view how the differences in macroscopic response are reflected in the spatial distributions of stress across the polycrystals. Here we primarily analyze the von Mises equivalent stress, a convenient scalar metric for quantifying the magnitude of the stress tensor. We present the distribution of von Mises stress across the gradient virtual microstructure (at the macroscopic loading points marked in Figure~\ref{fig:stress-strain}) in Figure~\ref{fig:von_Mises3d}a-d, and the same for the homogenized virtual microstructure in Figure~\ref{fig:von_Mises3d}e-h. As we compare the two distributions of von Mises stress, we note the presence of significant qualitative differences. In particular, we observe that the homogenized microstructure tends to exhibit smoother gradients along the stress field, while the spatial distribution of stresses in the gradient microstructure is more mottled, i.e., with more local peaks. However, the homogenized microstructure exhibits some of the largest stress peaks.

\begin{figure}[h!]
    \centering \includegraphics[width=1.0\textwidth]{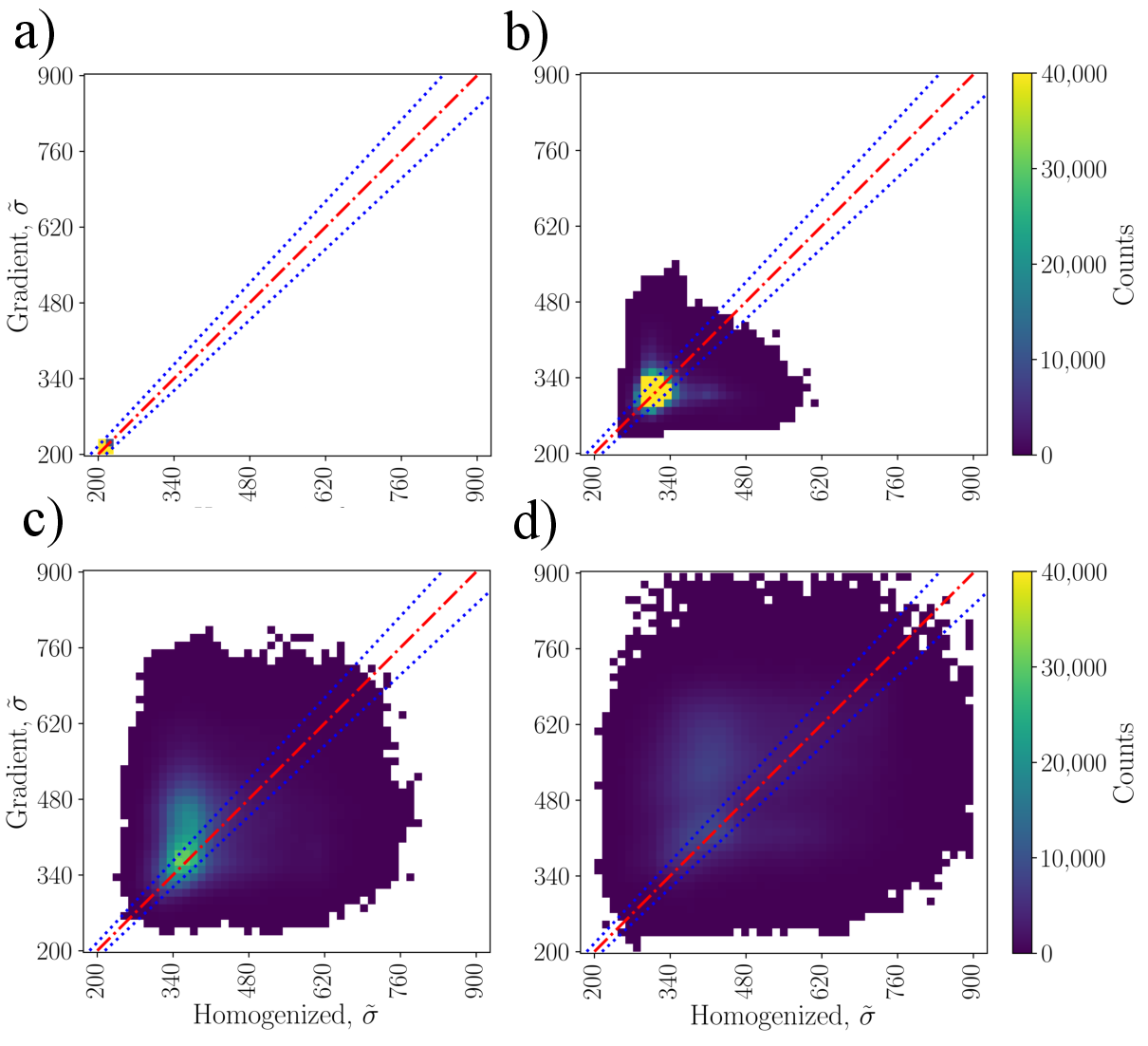}
    \caption{Two-dimensional histograms comparing the von Mises stresses, $\tilde{\sigma}$, between the homogenized and gradient virtual microstructures on an element-by-element basis. Histograms correspond to macroscopic engineering strains of a) 0.003, b) 0.011, c) 0.080, and d) 0.200. The dashed diagonal line indicates perfect correlation between the two virtual microstructures (i.e., 1-to-1), while the enveloping blue lines indicate a 10\% bandwidth (i.e., values that fall within $\pm$5\% of 1-to-1 correlation).} 
    \label{fig:von Mises_dist_hist}
\end{figure}

We can further view these differences between the predicted von Mises stresses on an element-by-element basis between the two polycrystals. We present this data in Figure~\ref{fig:von Mises_dist_hist}, which depicts two-dimensional histograms comparing the element-by-element von Mises stresses between the gradient and homogenized virtual microstructures at the same four macroscopic states of interest as before. For interpreting the figures, we note that the diagonal line indicates perfect correlation (``1-to-1'')---in other words, the histogram bins which fall on this line indicate the counts of elements which have the same von Mises stress in both the gradient simulation and the homogenized simulation. Histogram bins above the line indicate counts of elements which had a higher von Mises stress in the gradient sample, while those below the line indicate the opposite. White regions in the plots are bins that have no counts of elements which exhibit the given stress pairing between the gradient and homogenized microstructures. We observe, qualitatively, that as plasticity begins and evolves, the distribution of element counts tends to skew toward a positive deviation from the line of perfect correlation, indicating that more elements exhibit higher von Mises stresses in the gradient virtual microstructure compared to the homogenized microstructure. 

Turning specifically to Figure~\ref{fig:von Mises_dist_hist}a, we observe very little difference between the stresses in the elastic regime (i.e., the overwhelming majority of element counts are on or close to the 1-to-1 line). We do not find this entirely surprising as the elastic response of Al alloys is generally nearly isotropic (the elastic constants utilized in this study possess a Zener ratio of 0.94, where 1 is isotropic), so local changes in orientation are expected to have minimal effect on purely elastic behavior. With the onset of plasticity (Figure~\ref{fig:von Mises_dist_hist}b-d), we observe that the distributions of von Mises stresses begin to significantly deviate between the two samples, as evident by the appreciable spread in distribution of element counts away from the 1-to-1 line. Inspecting Figure~\ref{fig:von Mises_dist_hist}b, we note that this behavior is evident even shortly after macroscopic yield (i.e., with relatively minimal plastic deformation), where some elements exhibit von Mises stresses in either microstructure greater than \SI{200}{\mega\pascal} higher than in the other microstructure, which is orders of magnitude greater than the difference in macroscopic stresses of the two samples at this applied strain. This trend is increasingly clear at later states in plastic deformation, where at macroscopic engineering strains of 0.08 and 0.2, we calculate maximum differences on the order of \SI{400}{\mega\pascal} at the elemental scale between the two simulations (with the gradient microstructure increasingly exhibiting higher stresses than the homogenized microstructure). This result is, again, an order of magnitude greater than the magnitude of the differences observed in the macroscopic stresses. Additionally, we note that the peak of the histogram skews toward the gradient structure near a constant value for the homogenized structure, which overall indicates a relatively homogeneous stress field in the homogenized structure compared to a more varied (and indeed higher) stress field in the gradient structure.

To further quantify the trends observed in these plots, we gather statistics from the histograms in Figure~\ref{fig:von Mises_dist_hist}. In the figure, we plot the 1-to-1 correlation line, along with two lines representing a 10\% band, or an envelope containing values which fall within $\pm$5\% of 1-to-1. From the data, we can calculate the total number of elements that fall above the band (i.e., elements which have high-stress values in the gradient microstructure), the number of elements that fall within the band (i.e., elements which have values that fall ``within tolerance''), and the number of elements that fall below the band (i.e., elements which have high-stress values in the homogenized microstructure). Collectively, we thus define a metric for quantifying the approximate volume which skews towards more deviant states in either microstructure. We present these values in Table~\ref{tab:stress_hist_stats}. Overall, we observe that the stresses tend toward the extreme in the gradient microstructure more considerably than the homogenized microstructure at every strain state (save the first elastic state, where all elements fall within tolerance). This confirms what we observe qualitatively in Figure~\ref{fig:von Mises_dist_hist}.

\begin{table}[h!]
    \centering
    \begin{tabular}{c | c c c c }
    Histogram Region & $\bar{\varepsilon} = 0.003$ &$\bar{\varepsilon} = 0.011$ & $\bar{\varepsilon} = 0.080$ & $\bar{\varepsilon} = 0.200$ \\
    \hline
    Extreme Gradient & 0.00 & 18.28 & 50.11 & 57.90 \\
    Within Tolerance & 100.00 & 69.80 & 28.09 & 19.26 \\
    Extreme Homogenized & 0.00 & 11.92 & 21.80 & 22.84 \\
    \end{tabular}
    \caption{Statistics from the stress histograms detailing the percentage of the element counts with extreme gradient values (i.e., values where the gradient microstructure exhibit greater than 5\% above 1-to-1 correlation), within tolerance (i.e., values within $\pm$5\% of 1-to-1 correlation), and extreme homogenized values (i.e., values where the homogenized microstructure exhibits greater than 5\% above 1-to-1 correlation) at the four macroscopic engineering strain states of interest.}
    \label{tab:stress_hist_stats}
\end{table}

\subsection{Element-Scale Plastic Strain Behavior}

\begin{figure}[h!]
    \centering \includegraphics[width=1.0\textwidth]{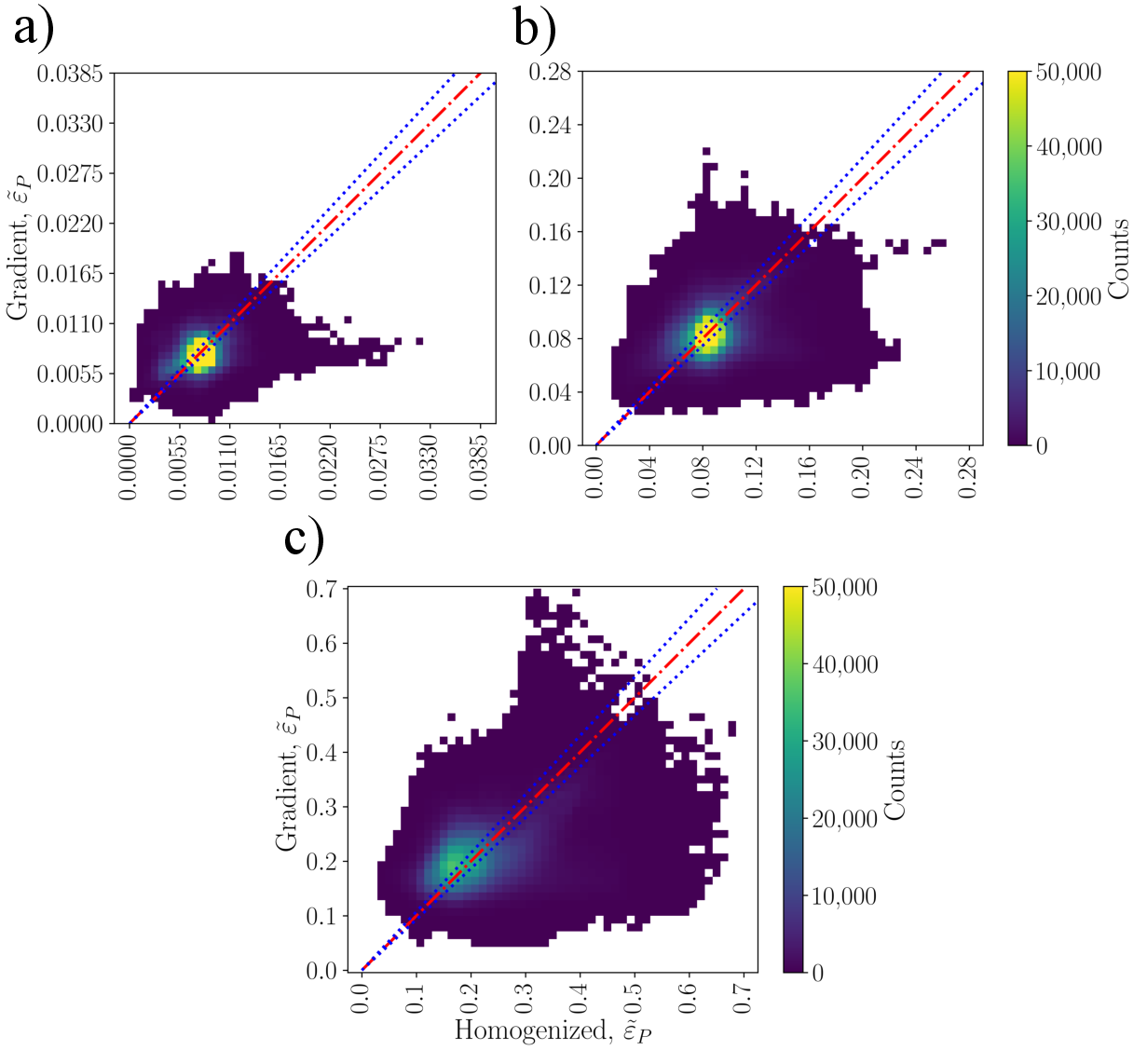}
    \caption{Two-dimensional histograms comparing the equivalent plastic strain, $\tilde{\varepsilon}_{P}$, between the homogenized and gradient virtual microstructures on an element-by-element basis. Histograms correspond to macroscopic engineering strains of a) 0.011, b) 0.080, and c) 0.200. The dashed diagonal line indicates perfect correlation between the two virtual microstructures (i.e., 1-to-1), while the enveloping blue lines indicate a 10\% bandwidth (i.e., values that fall within $\pm$5\% of 1-to-1 correlation).}
    \label{fig:epeq_dist_hist}
\end{figure}

We next inspect how the plastic strain behavior is affected by changes in initial orientation distributions between the two samples. Past studies have frequently established correlations between the localization of plasticity in a sample (and thus in crystals) and a sample's ultimate fatigue life~\cite{sangid2013physics}. Recently, the correlations were confirmed by a comprehensive study of strain localization across a wide range of alloy systems~\cite{stinville2022origins}, thus emphasizing the importance of inspecting the magnitude and distribution of plastic strain in polycrystalline materials. Similarly to Figure~\ref{fig:von Mises_dist_hist}, we depict two-dimensional histograms comparing the element-by-element equivalent plastic strain post yield between the gradient and homogenized microstructures in Figure~\ref{fig:epeq_dist_hist}. We observed no microplasticity in either microstructure at $\bar{\varepsilon}=0.003$ so the comparison of equivalent plastic strain is omitted. We note the difference of scale between Figure~\ref{fig:epeq_dist_hist}a-c compensating for the difference in total applied macroscopic strain. We observe that the plastic strain distributions tend to skew slightly toward the homogenized microstructure. In the most extreme cases, we observe that equivalent plastic strains in the homogenized microstructure are nearly 4 times of those in the gradient microstructure.

To further quantify the trends observed in these plots, we calculate the approximate volumes of the samples that tend to exhibit extreme values in either microstructure or fall within tolerance, similar to the analysis performed above on the stress distributions. We present these values in Table~\ref{tab:strainpl_hist_stats}. Here, we note that the homogenized microstructure generally tends to exhibit more extreme values than the gradient microstructure, except for in late plasticity where they exhibit nearly equal amounts of extreme values, confirming the previous qualitative observations. Collectively, this indicates that the homogenized microstructure must accommodate the deformation via a larger degree of local plastic deformation than the gradient microstructure, which overall agrees with the lower macroscopic hardening rate observed in the homogenized microstructure.

\begin{table}[h!]
    \centering
    \begin{tabular}{c | c c c }
    Histogram Region & $\bar{\varepsilon} = 0.011$ & $\bar{\varepsilon} = 0.080$ & $\bar{\varepsilon} = 0.200$ \\
    \hline
    Extreme Gradient & 32.37 & 35.23 & 41.54 \\
    Within Tolerance & 26.80 & 27.00 & 17.93 \\
    Extreme Homogenized & 40.84 & 37.77 & 40.53 \\
    \end{tabular}
    \caption{Statistics from the plastic strain histograms detailing the percentage of the element counts with extreme gradient values (i.e., values where the gradient microstructure exhibit greater than 5\% above 1-to-1 correlation), within tolerance (i.e., values within $\pm$5\% of 1-to-1 correlation), and extreme homogenized values (i.e., values where the homogenized microstructure exhibits greater than 5\% above 1-to-1 correlation) at the three macroscopic engineering strain states of interest.}
    \label{tab:strainpl_hist_stats}
\end{table}

\subsection{Slip Behavior}

\begin{figure}[h!]
    \centering \includegraphics[width=1.0\textwidth]{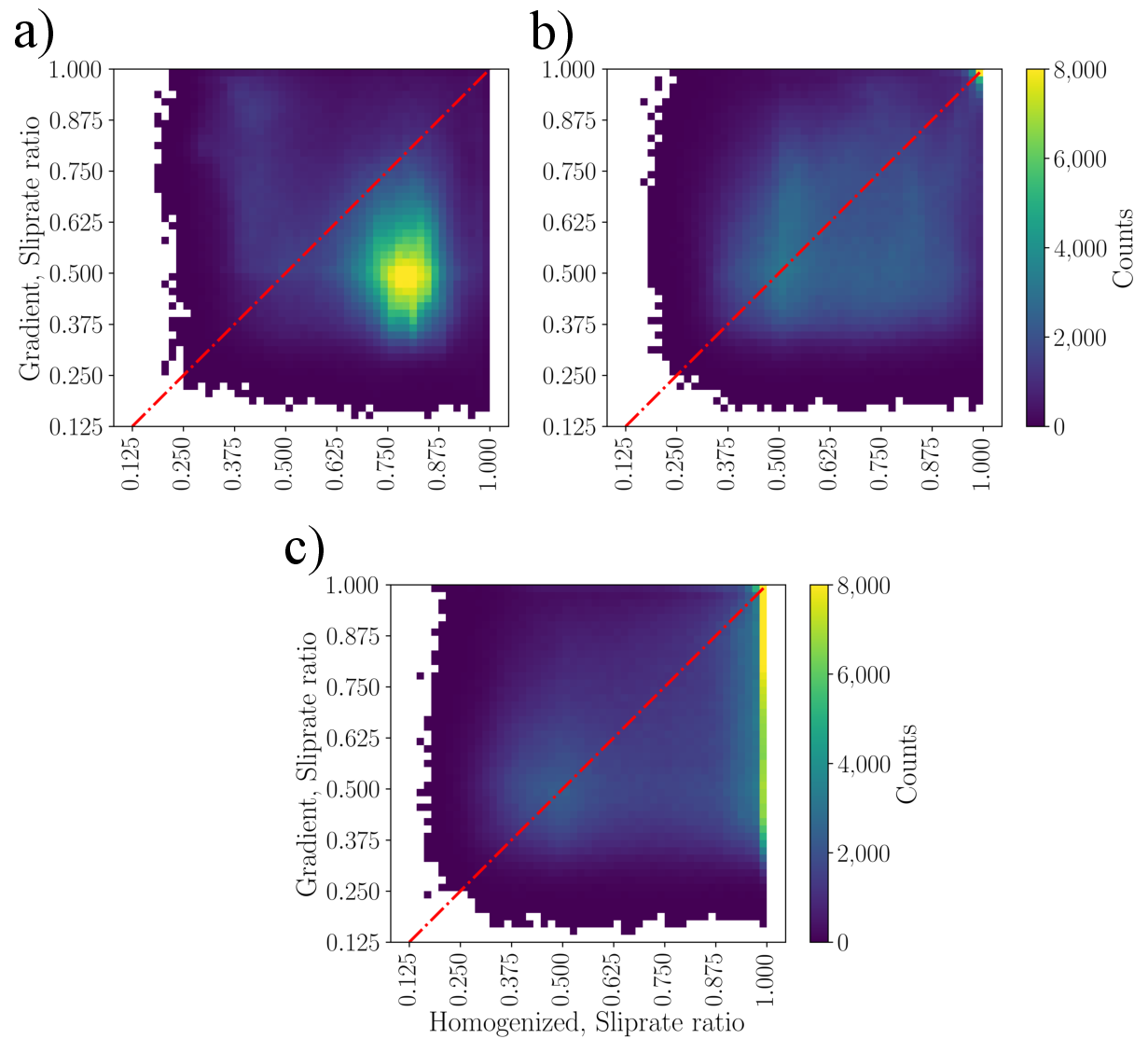}
    \caption{Two-dimensional histograms comparing the slip dominance ratio, $R$, between the homogenized and gradient virtual microstructures on an element-by-element basis. Histograms correspond to macroscopic engineering strains of a) 0.011, b) 0.080, and c) 0.200. The dashed diagonal line indicates perfect correlation between the two virtual microstructures (i.e., 1-to-1).}
    \label{fig:slip_ratios_mod_comp}
\end{figure}

Lastly, we wish to explore how the difference in the orientation distributions alters the active deformation modes, and in particular the character of slip. Here, we define a ratio $R$ as:
\begin{equation}
R=
\frac{
\max{|\dot{\gamma}_\alpha|}
}{
\sum_\alpha{|\dot{\gamma}_\alpha|}
}
\end{equation}
which provides a scalar metric of the degree to which slip activity is distributed among the 12 available slip systems. The extreme cases of this ratio are when slip activity is distributed equally among the 12 systems (i.e., a ratio 1/12, or 0.08), or conversely when only a single slip system is active (i.e., a ratio of 1). We compare these slip activity ratios between the two simulations in Figure~\ref{fig:slip_ratios_mod_comp} at the three loading points past the elastic-plastic transition. First examining Figure~\ref{fig:slip_ratios_mod_comp}a, we note that generally there exists higher $R$ ratios in the homogenized microstructure compared to the gradient microstructure. This indicates that the homogenized microstructure tends to exhibit dominant slip on a single, or limited, number of slip systems, compared to the gradient microstructure. In other words, we interpret this to indicate that the gradient microstructure generally relies on more slip systems for plastic deformation (indicated by lower $R$ ratios), compared to the homogenized microstructure. This is especially true at higher macroscopic strains, where the high counts in the histogram bins at the far right of Figures~\ref{fig:slip_ratios_mod_comp}b and c indicate that there exists a large volume fraction of the homogenized microstructure deforming purely by single slip, whereas there exists a larger spread of $R$ values along this line of bins in the gradient microstructure, indicating more varied slip.

\begin{figure}[h!]
    \centering \includegraphics[width=1.0\textwidth]{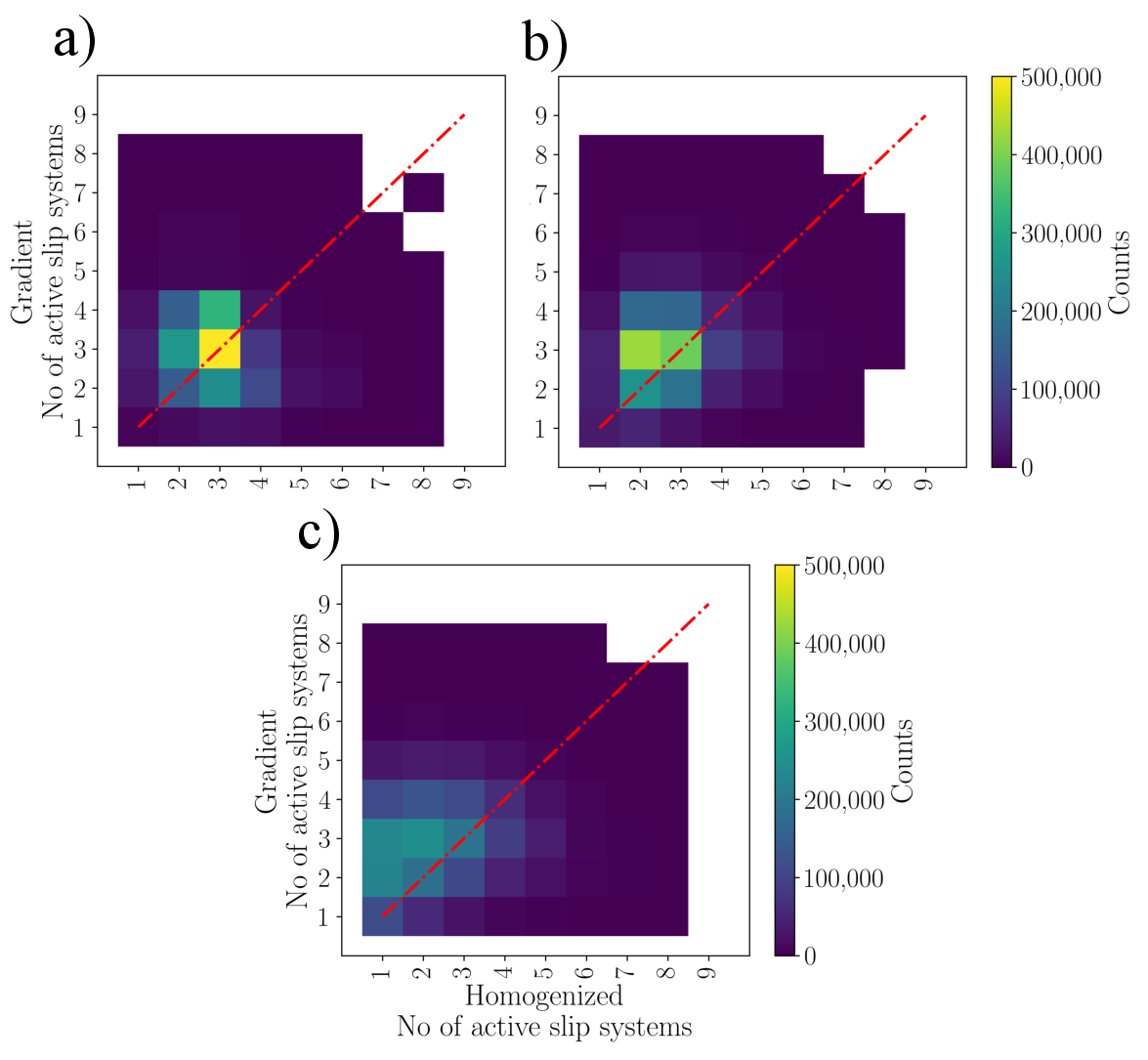}
    \caption{Two-dimensional histograms comparing active slip systems, defined as slip systems where $|\dot{\gamma}_\alpha|/\dot{\bar{\varepsilon}} > 0.1$, on an element by element basis. Histograms correspond to macroscopic engineering strains of a) 0.011, b) 0.080, and c) 0.200. The dashed diagonal line indicates perfect correlation between the two virtual microstructures (i.e., 1-to-1).}
    \label{fig:slip_activities}
\end{figure}

To provide further detail in this regard, we explore the number of active slip systems in either of the two simulations, the results of which are presented in Figure~\ref{fig:slip_activities}. For this analysis, we define a slip system being active as having a value of $|\dot{\gamma}_\alpha|$ greater than $0.1\dot{\bar{\varepsilon}}$ (i.e., the macroscopic applied strain rate). Overall, we observe that the gradient microstructure indeed tends to exhibit a more varied slip response compared to the homogenized microstructure. Again, this trend persists and becomes more apparent through the continued development of macroscopic plasticity---overall agreeing with the behaviors exhibited in Figure~\ref{fig:slip_ratios_mod_comp}. 

Collectively, we interpret this behavior in light of the forged sample's deformation history. The orientation gradients present in the sample developed due to the accommodation of plastic strain. Crystal plasticity theory dictates that the crystals will have (or would be) reoriented during deformation in an effort to reach a stress state in which multiple slip systems are active (i.e., toward a vertex of the single crystal yield surface~\cite{kumardawsonneoeulerian1998}). Upon deformation of the gradient microstructure, many crystal volumes are thus already aligned to accommodate plasticity on multiple slip systems. In the homogenized microstructure, conversely, by moving toward a homogenized granular orientation, we have reoriented portions of these crystals (potentially) away from the orientations to which they had evolved during thermomechanical processing to best accommodate the deformation processing. Thus, we should generally expect fewer active slip systems (at least initially) upon deforming the sample.

\section{Discussion}

Above, we present results examining variations in micromechanical response and slip system activation found when polycrystals contain intragranular orientation gradients that are often present in wrought alloys after thermomechanical processing. These simulations were made possible by novel HEDM reconstruction and mesh instantiation methodologies. We find that in the Al-7085 studied, the incorporation of intragranular orientation gradients leads directly to increased localization of stress, decreased localization of plastic strain, and increases in the number of slip systems activated to accommodate deformation. We interpret the differences in deformation response between these two microstructures as indication that neglecting intragranular orientation gradients in simulation instantiation (i.e., the typical practice) can lead to significantly different micromechanical responses from the true material being modeled. While there were only minor differences in the macroscopic stress-strain response (primarily hardening rates), the locations of extreme events (both stress and plastic deformation activity) shifted significantly between the two microstructure which in turn will lead to differences in failure locations and properties such as fatigue life.

We believe that neglect of these effects is likely not from lack of awareness, but rather that measurement of 3D orientation fields at sub-grain length scales has been historically out of reach. With maturation of 3D high-energy X-ray diffraction and 3D EBSD measurement capabilities (both of which now have commercial products available), consideration of 3D orientation fields for modeling purposes is increasingly feasible. Importantly, we believe that details of microstructural representation, such as intragranular orientation gradients, will be a key contributor to translating failure predictions from statistical to deterministic efforts. For the remainder of the discussion, we explore the generality of the modeling results presented in addition to implications for modeling of the fatigue life of alloy systems.

\subsection{Role of Elastic Anisotropy}

Aluminum alloys are relatively elastically isotropic, with Zener ratios close to 1. The Zener ratio for the elastic moduli used in all simulations presented thus far is 0.94, which led to very similar distributions of von Mises stress between the virtual microstructures in the elastic regime (see Figure~\ref{fig:von Mises_dist_hist}a), despite the difference in microtexture. To evaluate the role that elastic anisotropy would have on the micromechanical response, we perform a second simulation on the gradient microstructure, but with modified elastic properties delimited in the second row of Table~\ref{tab:elastic_properties}. Here, we modify the elastic moduli to increase the Zener ratio (to 2.98, or approximately 3$\times$ that of the original moduli) while maintaining the bulk modulus, which is mutually achieved via an increase in $C_{44}$. A Zener ratio near 3 is representative of a vast majority of FCC alloys including stainless steels, Cu alloys, and Ni-based superalloys~\cite{Bower2009}. We refer to simulations using the elastic moduli in the top and bottom rows of Table~\ref{tab:elastic_properties} as ``isotropic'' and ``anisotropic'' simulations, respectively. We note that the increase of shear modulus for the anisotropic moduli also has the effect of increasing the total stiffness of the microstructure.

\begin{figure}[h!]
      \centering \includegraphics[width=1.0\textwidth]{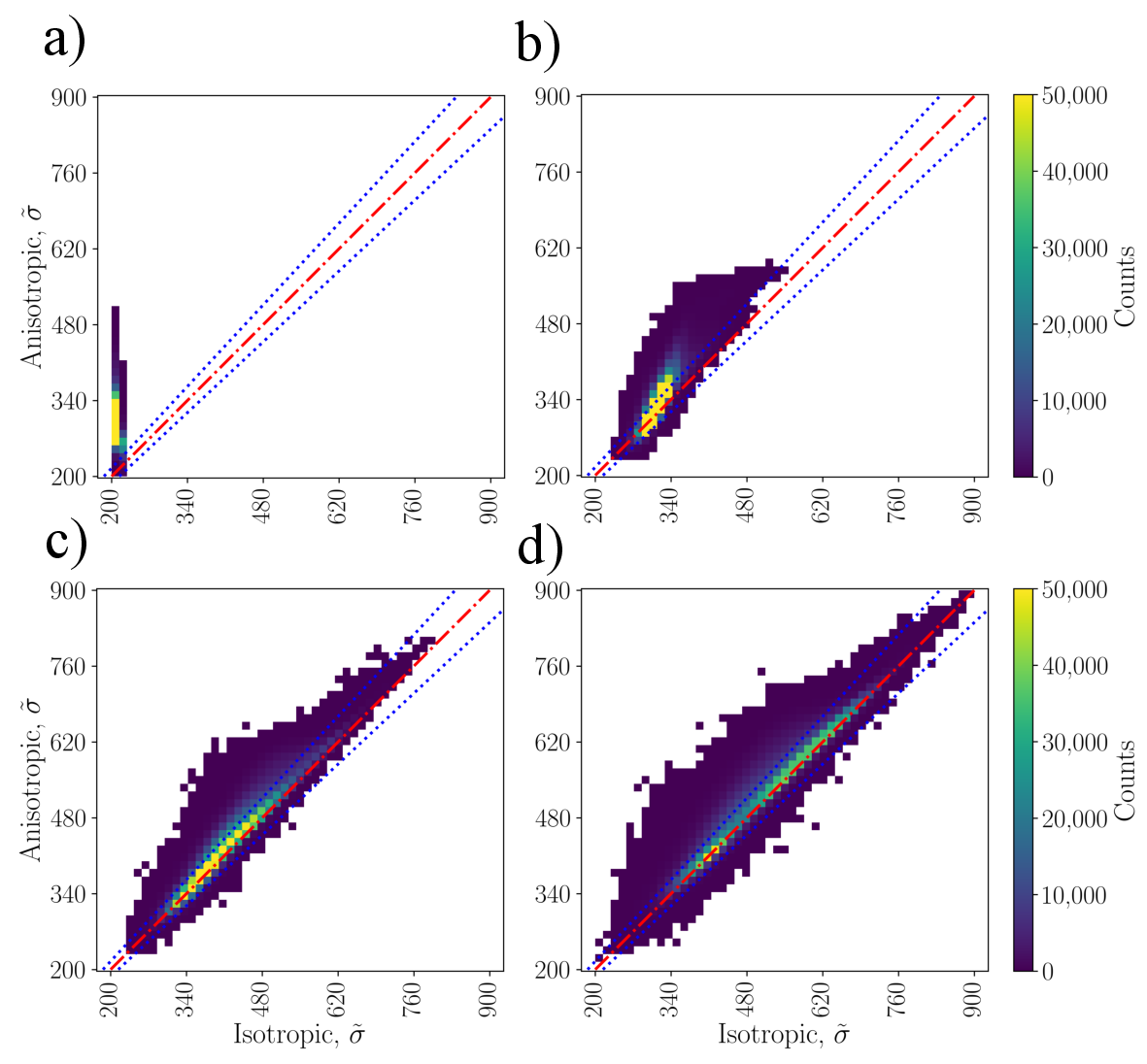}
    \caption{Two-dimensional histograms comparing the von Mises stresses, $\tilde{\sigma}$, between the gradient virtual microstructure using the ``isotropic'' and ``anisotropic'' elastic moduli sets on an element-by-element basis. Histograms correspond to macroscopic engineering strains of a) 0.003, b) 0.011, c) 0.080, and d) 0.200. The dashed diagonal line indicates perfect correlation between the two virtual microstructures (i.e., 1-to-1), while the enveloping blue lines indicate a 10\% bandwidth (i.e., values that fall within $\pm$5\% of 1-to-1 correlation).}
    \label{fig:eq_stress_mod_comp}
\end{figure}

We compare the element-by-element distributions of von Mises stress between the isotropic and anisotropic simulations at the four macroscopic strain points in Figure~\ref{fig:eq_stress_mod_comp}. We observe the largest differences between the anisotropic and isotropic simulations in the elastic regime, as shown in Figure~\ref{fig:eq_stress_mod_comp}a. The distribution is spread nearly vertically, which indicates that stress is nearly constant (at this macroscopic state) when employing nearly elastically isotropic moduli, but there is variation in stress state when employing elastically anisotropic moduli. Interestingly, once plasticity has fully developed (Figure~\ref{fig:eq_stress_mod_comp}c and~\ref{fig:eq_stress_mod_comp}d) we note that there is nearly perfect correlation between the stresses in the two microstructures. The behavior shortly after macroscopic yield (see Figure~\ref{fig:eq_stress_mod_comp}b) lies somewhere between these two extremes---the stresses between the anisotropic and isotropic simulations correlate better than in the fully elastic regime, but show more variation than when in fully-developed plasticity, indicating that the differences between the anisotropic and isotropic simulations quickly dissipate in the transitional region of macroscopic deformation. In total, we interpret the results as indicative of the fact that the anisotropy of elastic moduli will have a major effect on the elastic response prior to yield, but relatively minimal effect as plasticity becomes increasingly dominant.

\begin{figure}[h!]
      \centering \includegraphics[width=1.0\textwidth]{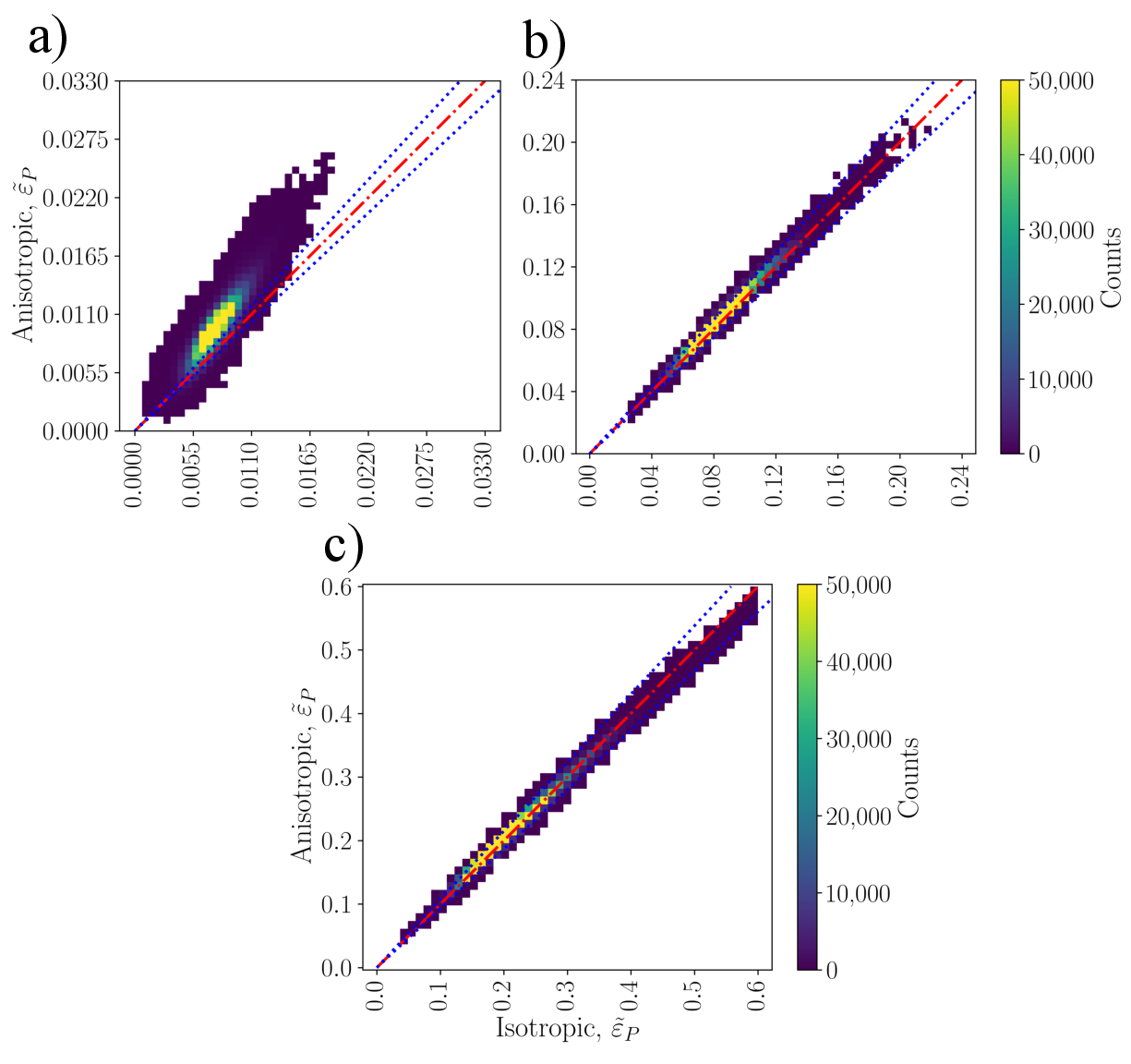}
    \caption{Two-dimensional histograms comparing the equivalent plastic strain, $\tilde{\varepsilon}_P$, between the gradient virtual microstructure using the ``isotropic'' and ``anisotropic'' elastic moduli sets on an element-by-element basis. Histograms correspond to macroscopic engineering strains past yield of a) 0.011, b) 0.080, and c) 0.200. The dashed diagonal line indicates perfect correlation between the two virtual microstructures (i.e., 1-to-1), while the enveloping blue lines indicate a 10\% bandwidth (i.e., values that fall within $\pm$5\% of 1-to-1 correlation).} 
    \label{fig:eq_pls_mod_comp}
\end{figure}

Similarly, we present Figure~\ref{fig:eq_pls_mod_comp}, which displays how altering the elastic anisotropy modifies the amount of local plasticity $\tilde{\varepsilon_P}$ (again driven by the von Mises stress) on an element-by-element basis. We make these comparisons only at macroscopic states where appreciable plastic deformation has occurred (again, the first state in the fully elastic regime is omitted). We note that the trends mirror those observed in the comparisons of the von Mises stress: the higher magnitude and spread of von Mises stresses drive more variability in plastic flow, but these effects are lost once the material has transitioned to fully developed plasticity.

\subsection{Effect of Orientation Gradients on Fatigue Indicator Parameters}

\begin{figure}[h!]
      \centering \includegraphics[width=1.0\textwidth]{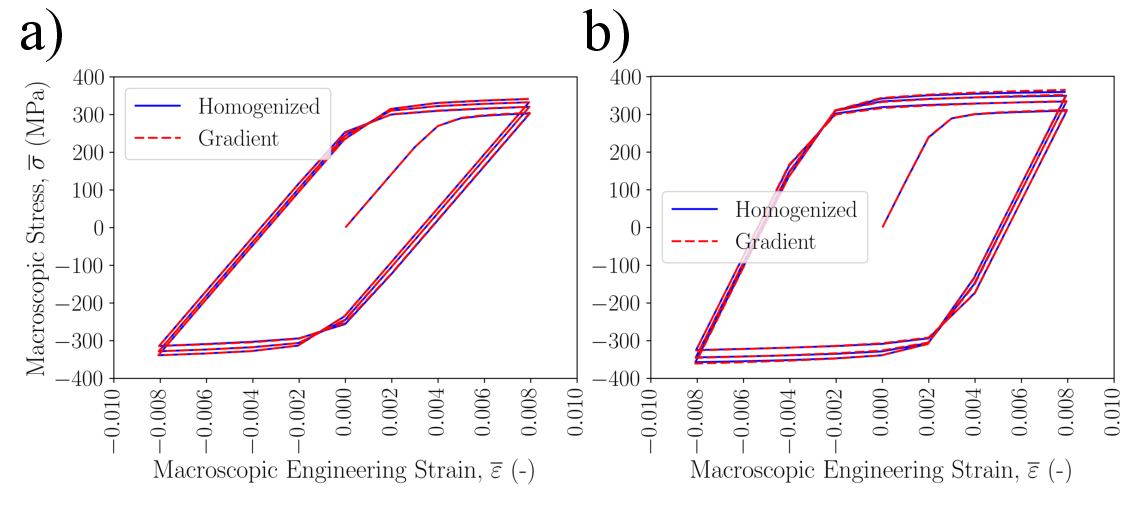}
    \caption{Macroscopic stress-strain curves for the homogenized and gradient virtual microstructures cyclically loaded for 3 cycles to macroscopic strain magnitudes of 0.008 using the a) ``isotropic'' and b) ``anisotropic'' elastic moduli sets.} 
    \label{fig:cyclic_load}
\end{figure}

As mentioned, the distributions of plastic deformation and stress, and more specifically the extreme values, influence the fatigue life of an alloy. To quantify these effects, different fatigue indicator parameters (FIPs) have been developed that incorporate measure of both local plastic strain and stress as a metric to predict the portions of polycrystalline materials which are more likely sites of the initiation of fatigue failure~\cite{mcdowell2010microstructure,przybyla2010microstructure,gu2023modeling}. Here, we primarily follow the approach developed by Stopka and co-workers~\cite{stopka2020microstructure,stopka2020microstructureb}, and in particular the FIP calculation approach found in studies performed in part on Al-7075-T6 (also a 7XXX aluminum alloy). In their approach, virtual specimens are cyclically loaded to fixed strain end points ($\tilde{\varepsilon}=\pm0.008$) for a small number of cycles (3) to ``shake-down'' the specimen (i.e., a number of cycles such that the majority of early-cyclic transience has diminished appreciably). From the final cycle of these simulations, a slip system $\mathrm{FIP}_\alpha$ is calculated, defined as:
\begin{equation}
\mathrm{FIP}_\alpha=\frac{\Delta \gamma_\alpha}{2}\left(1 + k \frac{\sigma_{n\alpha}}{\sigma_y} \right)
\end{equation}
where $\alpha$ in all cases denotes the slip system, $\Delta \gamma_\alpha$ is the difference in slip system shear strains at the cyclic end points, $\sigma_{n\alpha}$ is the maximum normal stress on a slip system, $\sigma_y$ is the macroscopic yield strength (see Table~\ref{tab:macro_response}), and $k$ is a material-dependent weighting of the normal stresses. The value of $k$ generally ranges from 0.5 to 1, but was found to be as high as 10 for Al-7075. Here, we follow this procedure and cyclically load both the homogenized and the gradient microstructures to macroscopic strain end points of $\pm0.008$ for 3 cycles, and follow the approach of Stopka et al. for the weighting of the normal stresses (i.e., $k=10$). We depict the macroscopic cyclic response of the homogenized (blue) and gradient (red dashed) microstructures in Figure~\ref{fig:cyclic_load}a. We observe cyclic hardening in the response of both microstructures, with the cyclic hardening being larger in the gradient microstructure---consistent with the increased hardening rate observed during the uniaxial loading.

\begin{figure}[h!]
      \centering \includegraphics[width=1.0\textwidth]{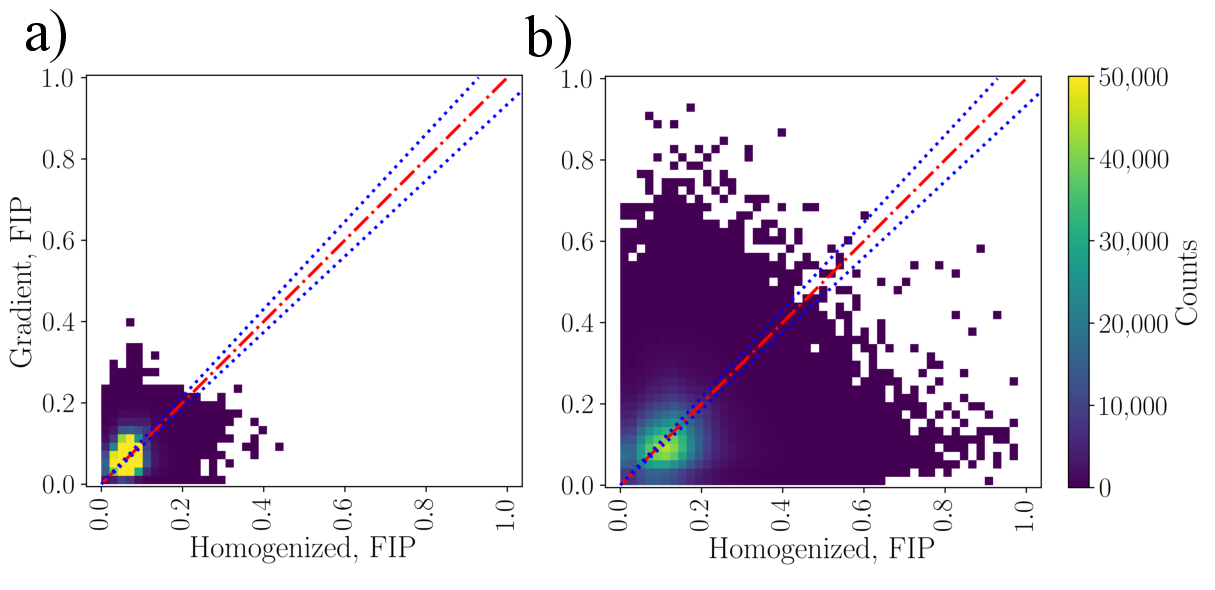}
    \caption{Two-dimensional histograms comparing the fatigue indicator parameters (FIPs) between the homogenized and gradient virtual microstructures on an element-by-element basis. Histograms correspond to the simulations performed with a) ``isotropic'' elastic moduli, and b)  ``anisotropic'' elastic moduli. The dashed diagonal line indicates perfect correlation between the two virtual microstructures (i.e., 1-to-1), while the enveloping blue lines indicate a 10\% bandwidth (i.e., values that fall within $\pm$5\% of 1-to-1 correlation).}
    \label{fig:fip}
\end{figure}

We determine the maximum $\mathrm{FIP}_\alpha$ in each element for these cyclic simulations. In support of the FIP calculations, we utilize the macroscopic yield strength as calculated by a 0.1\% offset method on the uniaxial simulation data. We present a comparison of the maximum $\mathrm{FIP}_\alpha$ in the elements between the two samples in Figure~\ref{fig:fip}a. Here, we observe that the $\mathrm{FIP}_\alpha$ values tend to be larger in the homogenized microstructure compared to those in the gradient microstructure. Specifically, we calculate FIP values in the homogenized microstructure can be nearly five times as high as in the same elements in the gradient microstructure (and, conversely, some elements nearly four times higher in the gradient microstructure compared to the homogenized microstructure). Consequently---and importantly regarding the comparison of the two microstructural instantiation methods considered here---we would expect a corresponding decrease in fatigue life of the homogenized microstructure. This observation is not entirely surprising, considering the homogenized microstructure's general propensity for regions of higher plasticity, as described above. However, it emphasizes that intragranular orientation gradients are a critical feature that (when plausibly available) must be incorporated into virtual microstructures for more accurate predictions. Similar to the analyses found in the results, we can quantify the approximate volumes of material exhibiting extreme FIP values skewed toward either microstructure or those which fall within tolerance. We present these values in Table~\ref{tab:fip_hist_stats}. Here, we note that the largest volume of material exhibits FIP values which tend toward the extreme in the gradient microstructure, indicating that larger volume fraction of the gradient microstructure will display more propensity for fatigue failure. However, since fatigue initiates due to local phenomena, we cannot ignore the fact that the homogenized microstructure displays the highest overall FIP value between the two microstructures, as described above.

\begin{table}[h!]
    \centering
    \begin{tabular}{c | c c}
    Histogram Region & Isotropic & Anisotropic \\
    \hline
    Extreme Gradient & 46.32 & 42.15 \\
    Within Tolerance & 11.97 & 7.96 \\
    Extreme Homogenized &  41.71 & 49.89\\
    \end{tabular}
    \caption{Statistics from the FIP histograms detailing the percentage of the element counts with extreme gradient values (i.e., values where the gradient microstructure exhibit greater than 5\% above 1-to-1 correlation), within tolerance (i.e., values within $\pm$5\% of 1-to-1 correlation), and extreme homogenized values (i.e., values where the homogenized microstructure exhibits greater than 5\% above 1-to-1 correlation) for both the simulations performed with ``isotropic'' and ``anisotropic'' elastic moduli.}
    \label{tab:fip_hist_stats}
\end{table}

As discussed in the previous subsection, anisotropy of elastic response has the largest effect in the elastic regime and in the elasto-plastic transition. As such, it is plausible that elastic anisotropy will play a role during cyclic loading, assuming the strain states are not appreciably large so as to diminish the influence of elastic anisotropy, as demonstrated previously. For this reason, we repeat the cyclic loading simulations and FIP analysis for the homogenized and gradient microstructures, but using the previously-introduced anisotropic elastic moduli. We present the macroscopic cyclic response for the gradient and homogenized microstructures in Figure~\ref{fig:cyclic_load}b, as well as the element-by-element comparison of the maximum slip system $\mathrm{FIP}_\alpha$ in Figure~\ref{fig:fip}b. In both microstructures, we observe a moderate increase in the amount of cyclic hardening when considering the anisotropic moduli. Additionally, with increased anisotropy (and stiffness), we note that the maximum $\mathrm{FIP}_\alpha$ values in both microstructures increases (in addition to the previously-observed trend of the gradient structure exhibiting higher $\mathrm{FIP}_\alpha$ values). From these $\mathrm{FIP}_\alpha$ comparisons, we interpret the trends to indicate that increased elastic anisotropy magnifies the effects of intragranular orientation gradients, and will further lower fatigue life in comparison to an elastically isotropic alloy. However, we note that there is less difference in the histograms for the anisotropic simulations compared to the isotropic simulations, i.e., both simulations display maximum FIP values of similar magnitude, indicating their maximum likelihood of fatigue to be similar. Statistics from the histograms are again presented in Table~\ref{tab:fip_hist_stats}, which here, instead, show that the homogenized microstructure now has a higher volume fraction with extreme FIP values. In light of the similar maximum FIP values in both microstructures, we now give more weight to the difference in volume fractions to indicate fatigue performance, which thus again indicates that the homogenized microstructure would fail earlier than the gradient microstructure. 

We note that this behavior, in part, can be explained by the overall increase in macroscopic elastic modulus (a consequence of the way in which we increased elastic anisotropy while holding bulk modulus fixed), as seen in Figure~\ref{fig:cyclic_load}. The earlier onset of plasticity, as caused by the stiffer elastic response but fixed initial slip system strength, leads to a slight increase in effective plastic strain values in the anisotropic samples. This is further to be expected, as we perform the cyclic simulations to strains in the early plastic regime where we expect slightly higher plastic strain values in the anisotropic samples (see Figure~\ref{fig:eq_pls_mod_comp}a for similar behavior in the monotonic simulations at similar strain states), compared to later plasticity, where the differences are muted.

\section{Summary}

In this study, we have presented a novel method to reconstruct 3D orientation fields with large amounts of intragranular misorientation. We demonstrate the approach through reconstruction of a forged Al-7085 specimen---containing both low-angle boundaries and large amounts of intragranular misorientation ($>10^\circ$)--- and utilize the reconstructed sample to explore the role of intragranular orientation heterogeneity on both macroscopic and microscopic mechanical deformation response. We utilized crystal plasticity finite element simulations to predict the deformation response of these two microstructures subjected to both monotonic and cyclic loading, and additionally altered the elastic parameters to explore the influence of elastic anisotropy on the observed trends as an effort to lend insight to expected behavior in other material systems. From the simulation results, we found that---relative to a microstructure with homogenized grain orientations---the incorporation of intragranular orientation gradients:
\begin{enumerate}
\item increases the macroscopic hardening rate,
\item produces higher stresses, lower degrees of plastic strain, and more varied slip response,
\item and will likely have a higher fatigue life, as informed via the presence of lower FIP values.
\end{enumerate}
We found that these effects are exacerbated when the anisotropy of the elastic response is increased. While the generality of these findings across different alloy classes and microstructures deserves further exploration, we contend that it is clear that intragranular orientation gradients can significantly alter the prediction of the mechanical response, and should be considered whenever they are present in the material of interest.

\section*{Data Availability Statement}

The raw/processed data required to reproduce these findings cannot be shared publicly at this time due to technical or time limitations. Data can be made available upon reasonable request.

\section*{Acknowledgments}

This work is based on research conducted at the Center for High-Energy X-ray Sciences (CHEXS), which is supported by the National Science Foundation (BIO, ENG and MPS Directorates) under award DMR-1829070. KS was funded through the duration of this study by The Boeing Company as part of its Learning Together Program, and further received resources from the University of Alabama. MPK received support and resources from the University of Alabama during this study. MS and DCP received support from the Penn State Department of Materials Science and Engineering.

\bibliographystyle{elsarticle-num}
\bibliography{References.bib}

\end{document}